\documentclass[onecolumn,twoside,draftclsnofoot]{IEEEtran}
\usepackage{cite}
\usepackage{CJK}
\usepackage[cmex10]{amsmath}
\usepackage{amsfonts}
\usepackage{amssymb}
\usepackage{paralist}
\usepackage[dvips]{graphicx}
\usepackage[caption=false,font=footnotesize]{subfig}
\usepackage{booktabs}
\usepackage{bm}
\usepackage{algorithm}
\usepackage{algorithmicx}
\usepackage{algpseudocode}
\usepackage{multicol}
\usepackage{stfloats}
\ifCLASSINFOpdf
\else
\fi
\begin{document}
%
\title{An Overview of Transmission Theory and Techniques of Large-scale Antenna Systems for 5G Wireless Communications}
%
%
%

\author{Dongming~Wang,
        Yu~Zhang,
        Hao~Wei,
        Xiaohu~You,~\IEEEmembership{Fellow,~IEEE,}
        Xiqi~Gao,~\IEEEmembership{Fellow,~IEEE,}
        and~Jiangzhou~Wang,~\IEEEmembership{Senior Member,~IEEE}
\thanks{D. Wang, Y. Zhang, H. Wei, X. You and X. Gao are with National Mobile Communications Research  Laboratory, Southeast University, Nanjing 210096, China (Email:\{wangdm, yuzhang, weihao, xhyu, xqgao\}@seu.edu.cn).} \par
\thanks{J. Wang is with the School of Engineering and Digital Arts, University of Kent, Canterbury, Kent, CT2 7NZ, U.K (Email:j.z.wang@kent.ac.uk).} \par
}

\maketitle

\begin{abstract}
To meet the future demand for huge traffic volume of  wireless data service, the research on the fifth generation (5G) mobile communication systems has been undertaken in recent years. It is expected that the spectral and energy efficiencies in 5G mobile communication systems should be ten-fold higher than the ones in the fourth generation (4G) mobile communication systems. Therefore, it is important to further exploit the potential of spatial multiplexing of multiple antennas. In the last twenty years, multiple-input multiple-output (MIMO) antenna techniques have been considered as the key techniques to increase the capacity of wireless communication systems. When a large-scale antenna array (which is also called massive MIMO) is equipped in a base-station, or a large number of distributed antennas (which is also called large-scale distributed MIMO) are deployed, the spectral and energy efficiencies can be further improved by using spatial domain multiple access. This paper provides an overview of massive MIMO and large-scale distributed MIMO systems, including spectral efficiency analysis, channel state information (CSI) acquisition, wireless transmission technology, and resource allocation.
\end{abstract}


%
\IEEEpeerreviewmaketitle

\section{Introduction}
In recent years, mobile data traffic has been increasing almost exponentially. According to the most recent prediction from Ericsson, global mobile data traffic will increase nearly elevenfold from 2015 to 2020~\cite{Ericsson}. At the same time, as the energy consumption of information and communications technology becomes large, it is very urgent to reduce the energy consumption of mobile communication systems. This definitely leads to a great challenge for current fourth generation (4G) mobile communications, and also provide a big chance for the fifth generation (5G) mobile communication systems.

In order to meet the future demand of the mobile data service, one of  the fundamental objectives of 5G mobile communication systems is to further improve the spectral and energy efficiencies by one order of magnitude higher than the ones in 4G mobile communication systems. This requires new revolution of network architecture and wireless transmission technologies, to fundamentally address the problem of the spectral and energy efficiencies of mobile communication systems, as well as achieve the goals of higher data rate and green wireless communications~\cite{you2014}\cite{Ma}.

Multiple-input multiple-output (MIMO) antennas, as a breakthrough communication technique in the past 20 years, is a fundamental approach to exploit spatial domain resource. MIMO offers diversity gain, multiplexing gain and power gain~\cite{Paulraj}, to improve the  reliability, support the spatial multiplexing of both single and multiple users, and increase the energy efficiency through  beamforming techniques, respectively. So far, MIMO technology has been adopted by third generation partnership project (3GPP) long-term evolution (LTE),  IEEE 802.11ac and other wireless communication standards. However, for 4G mobile communication systems, as only a small number of antennas is equipped in base stations (BSs)~\cite{Marzetta_WC}, the spatial resolution is limited and then the performance gain is not fully exploited. Furthermore, under the current system configuration, the implementation of the capacity-approaching transmission scheme is extremely difficult.

Distributed antenna system (DAS) is another approach to exploit spatial dimension resources~\cite{You_X_H_WC,Zhu2011,JWang2012,Osman}. In DAS, geographically dispersed Remote Radio Units (RRUs) are equipped with multiple antennas, and are connected to a baseband unit (BBU) through high-speed backhaul links. Similar to MIMO, with the cooperation between RRUs, DAS can  serve single or multiple mobile terminals in the same time-frequency resource. Then, it is also called as distributed MIMO system or cooperative MIMO system. Distributed MIMO technology could not only obtain three types of gains of MIMO, but also get the macro-diversity and the power gain due to smaller path loss~\cite{Dai2011A,Wang2013Asymptotic,WangJSAC}.

\begin{figure}[!t]
\centering
\includegraphics[scale=.68]{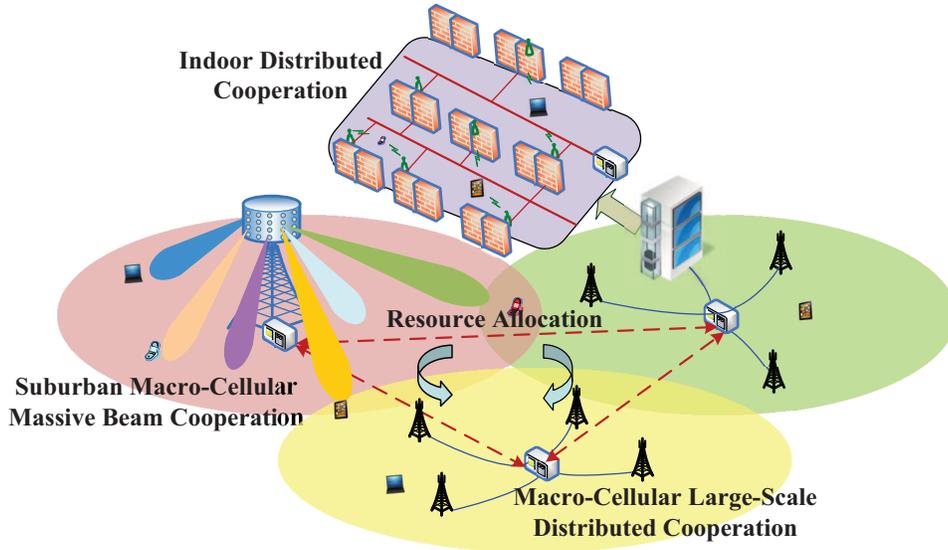}
\caption{Large-scale cooperative wireless communications.}
\label{fig1}
\end{figure}

To further improve the spectral and energy efficiencies of 4G system, both industry and academia have reached a consensus to increase the number of cooperative RRUs at hot spots ~\cite{Huh_WC, wang2015Large}, or replace the current multiple antennas with a large-scale antenna array in each BS ~\cite{Marzetta_WC}. Both the technologies can be viewed as a large-scale collaborative wireless communication shown in Figure~\ref{fig1}.

As shown in Figure~\ref{fig1}, a large number of antennas could be dispersed within a cell (called large-scale distributed MIMO), or centrally deployed at a BS (referred to as massive MIMO). Theoretical research and preliminary performance assessments ~\cite{Marzetta_WC, wang2015Large} demonstrated that as the number of BS antennas (or the number of distributed RRUs) becomes infinity, inter-user channels will become orthogonal. In this case, Gaussian noise and inter-cell interference from other cells will be averaged out to zero and the transmit power of any user can be arbitrarily low. The system capacity is only limited by the reuse of the pilot resource.

In both massive MIMO and large-scale distributed MIMO systems, there exists a  consensus on exploitation of wireless spatial dimension resources.  That is,  they have similar problems in the performance analysis and system design. Thus, in this paper, if not specifically stated, both massive MIMO and large-scale distributed MIMO systems are referred to as large-scale antenna systems (LSAS). Different from traditional MIMO system, LSAS introduces the challenges on both theoretical analysis and system design, which include spectral efficiency analysis under the constrain of pilot resource, channel state information (CSI) acquisition, wireless transmission technology, resource allocation, etc..

In the following, we will present the state of the art in LSAS. Firstly, we introduce its spectral efficiency analysis with imperfect CSI. Secondly, its bottleneck technology, CSI acquisition techniques including pilot design, channel estimation and reciprocity acquisition methods, is highlighted. Then, its transmission technology, consisting of multi-user precoding and joint multi-user detection, are reviewed. Finally, some resource allocation methods are introduced.

The symbols used in this paper are defined as follows. All boldface italic letters stand for vectors (lower case) or matrices (upper case). ${\bm{I}}$ denotes any suitable unit matrix. ${\left[ {\bm{A}} \right]_{ij}}$ represents the $i$th row and $j$th column element of the matrix ${\bm{A}}$. The superscripts ${\left(  \cdot  \right)^*}$, ${\left(  \cdot  \right)^{\rm{T}}}$ and ${\left(  \cdot  \right)^{\rm{H}}}$ respectively denote the matrix transpose, conjugate and conjugate transpose. The operator $\otimes$ denotes the Kronecker product. vec($\bm{A}$) denotes vectorization of a matrix. ${\rm{Tr}}\left( {\bm{A}} \right)$ and $\det \left( {\bm{A}} \right)$ stand for determinant and trace, respectively. The operator ${\text{diag}}\left( {\bm{x}} \right)$ denotes the diagonal matrix with ${\bm{x}}$ along its main diagonal. A block diagonal matrix denoted as ${\rm{diag}}\left[ {\begin{array}{*{20}{c}}{{{\bm{A}}_1}}& \cdots &{{{\bm{A}}_n}}\end{array}} \right]$ has blocks ${\bm{A}_1},\; \cdots \;,{\bm{A}_n}$ along its main diagnal. The operator ${\cal E}\left(  \cdot  \right)$ denotes expectation and the covariance operator is given by ${\mathop{\rm cov}} \left( {\bm{x},\bm{y}} \right) \buildrel \Delta \over = {\cal E}\left( {\bm{x}{\bm{y}^{\rm{H}}}} \right)\, - {\cal E}\left( \bm{x} \right){\cal E}\left( {{\bm{y}^{\rm{H}}}} \right)$. ${\cal C}{\cal N}\left( {\mu ,{\sigma ^2}} \right)$ stands for complex Gaussian distribution with mean $\mu$ and variance ${\sigma ^2}$.

\section{Spectral efficiency analysis of LSAS}
Capacity analysis is a basis for system design. In the past 20 years, many researchers have studied the capacity of MIMO. Random matrix theory plays an important role in the theoretical analysis of MIMO capacity. The fundamental mathematical tools include the statistical properties of Wishart matrix and its eigenvalues~\cite{Tulino}, the asymptotic statistical properties of large dimensional random matrices ~\cite{Tulino}, free probability theory~\cite{Luanan} and the deterministic equivalent method ~\cite{zhang_capacity_2013}. With these mathematical tools, extensive research has been done on the capacity of different MIMO channels, including multi-user MIMO (MU-MIMO) and distributed MU-MIMO channels.

However, most of the studies on MIMO capacity are focused on perfect CSI at receiver side. Since a bottleneck of multi-user large-scale MIMO systems is CSI acquisition, it is necessary to study information theory under the constrained pilot resource. The spectral and energy efficiencies in both uplink and downlink massive MIMO systems were analyzed in~\cite{Ngo}, where the impact of imperfect CSI on the performance was demonstrated. In~\cite{Hoydis}, the spectral efficiency of massive MIMO systems with different transceivers has been studied with the consideration of pilot contamination. The theoretical results revealed the relationship between the number of BS antennas per user and the transceiver techniques.

In the following, we consider multi-cell multi-user DAS as an example, of which massive MIMO is its special case. We assume that the pilot resource is limited and pilot sequences are reused by the users in the systems. We present a closed-form expression of asymptotic capacity. Furthermore, we also give some research directions.

\subsection{Theoretical analysis of spectral efficiency}
\begin{figure}[!t]
\centering
\includegraphics[scale=1]{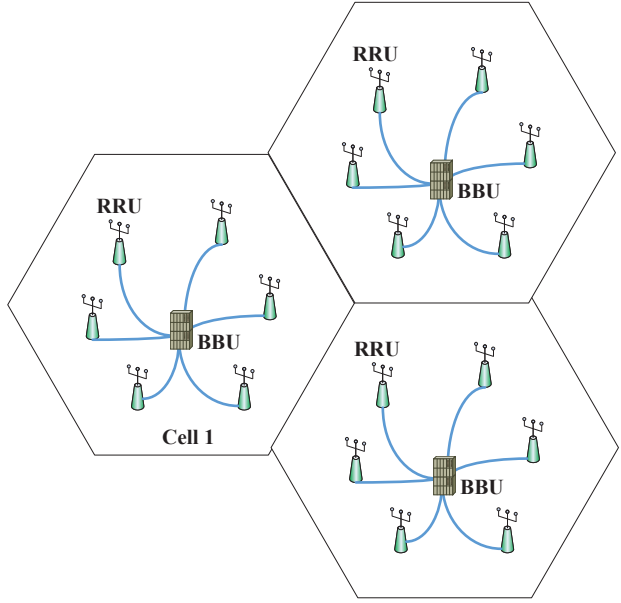}
\caption{Multi-cell multi-user large-scale distributed antenna system.}
\label{fig2}
\end{figure}

 Consider  multi-cell multi-user DAS with $L$ hexagonal cells, as shown in Figure~\ref{fig2}. Each cell consists of $N$ RRUs and $K$ single-antenna users. Each RRU is equipped with $M$ antennas. All the RRUs in each cell are connected to a BBU through optical fibers. In this paper, Cell 1 is the reference cell. The uplink channel vector between the $k$th user of cell $l$ to all of the RRUs of the cell 1 is denoted as
\begin{equation}
{{\bm{g}}_{l,k}} \buildrel \Delta \over = {\bm{R}}_{l,k}^{\frac{1}{2}}{\bm{\Lambda }}_{l,k}^{\frac{1}{2}}{{\bm{h}}_{l,k}},
\label{eq1}
\end{equation}
where
\begin{eqnarray}
\label{eq2}
& {{\bm{\Lambda '}}_{l,k}} \buildrel \Delta \over = {\rm{diag}}\left[ {\begin{array}{*{20}{c}}
{{\lambda _{l,1,k}}}& \cdots &{{\lambda _{l,N,k}}},\end{array}} \right],\\
\label{eq3}
& {{\bm{\Lambda }}_{l,k}} \buildrel \Delta \over = {{\bm{\Lambda '}}_{l,k}} \otimes {{\bm{I}}_M},\\
\label{eq4}
& {{\bm{R}}_{l,k}} \buildrel \Delta \over = {\rm{diag}}\left[ {\begin{array}{*{20}{c}}
{{{\bm{R}}_{l,1,k}}}& \cdots &{{{\bm{R}}_{l,N,k}}}\end{array}} \right],\\
\label{eq5}
& {{\bm{h}}_{l,k}}{\rm{ = }}{\left[ {\begin{array}{*{20}{c}}
{{\bm{h}}_{l,1,k}^{\rm{T}}}& \cdots &{{\bm{h}}_{l,N,k}^{\rm{T}}}\end{array}} \right]^{\rm{T}}}.
\end{eqnarray}
${\lambda _{l,n,k}}$ represents the large-scale and shadow fading between the $k$th user of cell $l$ and the $n$th RRU of cell 1 and ${{\bm{\Lambda '}}_{l,k}}$ is a diagonal matrix representing the large-scale and shadow fading matrix between the $k$th user of cell $l$ and all the RRUs of cell 1. We assume the links between the user and all the antennas of a RRU have the same large-scale and shadow fading and ${{\bm{\Lambda}}_{l,k}}$ represents the large-scale and shadow fading matrix between the $k$th user of cell $l$ and all the antennas of $N$ RRUs of cell 1. ${{\bm{R}}_{l,n,k}}$ is an $M \times M$ receive correlation matrix of the $k$th user of cell $l$ to the $n$th RRU of cell 1 and ${{\bm{R}}_{l,k}}$ is an $MN \times MN$ block diagonal receive correlation matrix of the $k$th user of cell $l$ to all the RRUs of cell 1. ${{\bm{h}}_{l,n,k}}$ represents the small-scale fading between the $k$th user of cell $l$ and the $n$th RRU of cell 1, and it is a vector of size $M$ which contains independent identically distributed (i.i.d.) zero mean circularly symmetric complex Gaussian (ZMCSCG) random variables with unit variance. ${{\bm{h}}_{l,k}}$, an $MN \times 1$ vector, represents the small-scale fading between the $k$th user of cell $l$ and all the antennas of $N$ RRUs of cell 1. The channel matrix between $K$ users of cell $l$ and all antennas of $N$ RRUs of cell 1 is defined as ${{\bm{G}}_l} = \left[ {\begin{array}{*{20}{c}} {{{\bm{g}}_{l,1}}}& \cdots &{{{\bm{g}}_{l,K}}} \end{array}} \right].$

We assume the BBU obtains uplink CSI by uplink pilot sequences from the users. $K \times K$ identity matrix is used as the pilot sequences set for $K$ users of any cell. That is,  for inner-cell users, the pilot sequences are orthogonal. For example, a user sends pilot symbol 1 on a time-frequency resource when other users in the same cell send pilot symbol 0. Since the pilot sequences are reused by the other cell users, there is a severe pilot contamination among the adjacent cells.

Because ${\cal E}\left[ {{{\bm{g}}_{l,k}}{\bm{g}}_{l',k'}^{\rm{H}}} \right] = 0$, for $k \ne k'$, the estimation of ${{\bm{g}}_{l,k}}$ can be processed individually. To estimate ${{\bm{g}}_{l,k}}$, we have the following observations
\begin{equation}
{{\bm{y}}_{{\rm{P}},k}} = {{\bm{g}}_{l,k}} + \sum\limits_{i \ne l} {{{\bm{g}}_{i,k}}}  + {{\bm{z}}_{{\rm{P}},k}},
\label{equ6}
\end{equation}
where ${{\bm{z}}_{{\rm{P}},k}}$ is the equivalent Gaussian noise with variance ${\gamma _{\rm{P}}}$. From \cite{wang2013ICC}, we can conclude that
\begin{eqnarray}
\label{equ7}
& {{\bm{\hat g}}_{l,k}} = {{\bm{R}}_{l,k}}{{\bm{\Lambda }}_{l,k}}{\bm{Q}}_k^{ - 1}{{\bm{y}}_{{\rm{P}},k}},\;\;l = 1, \ldots ,L.\\
\label{equ8}
& {{\bm{Q}}_k} = \sum\limits_{i = 1}^L {{{\bm{R}}_{i,k}}{{\bm{\Lambda }}_{i,k}}}  + {\gamma _{\rm{P}}}{{\bm{I}}_{MN}}.
\end{eqnarray}
Channel estimation error is defined as ${{\bm{\tilde g}}_{l,k}} = {{\bm{g}}_{l,k}} - {{\bm{\hat g}}_{l,k}}$ and its covariance matrix can be expressed as
\begin{equation}
{\mathop{\rm cov}} \left( {{{{\bm{\tilde g}}}_{l,k}},{{{\bm{\tilde g}}}_{l,k}}} \right) = {{\bm{R}}_{l,k}}{{\bm{\Lambda }}_{l,k}} - {{\bm{R}}_{l,k}}{{\bm{\Lambda }}_{l,k}}{\bm{Q}}_k^{ - 1}{{\bm{R}}_{l,k}}{{\bm{\Lambda }}_{l,k}}.
\label{equ9}
\end{equation}
Define ${{\bm{\hat h}}_k} \triangleq {\bm{Q}}_k^{ - \frac{1}{2}}{{\bm{y}}_{{\rm{P}},k}}$. It can be seen that ${{\bm{\hat h}}_k}$ obeys ${{\bm{\hat h}}_k} \sim {\cal C}{\cal N}\left( {{\bm{0}},{{\bm{I}}_{MN}}} \right)$ and is not related to $l$. Then, the channel vector can be modelled as
\begin{equation}
{{\bm{\hat g}}_{l,k}} = {{\bm{R}}_{l,k}}{{\bm{\Lambda }}_{l,k}}{\bm{Q}}_k^{ - \frac{1}{2}}{{\bm{\hat h}}_k}.
\label{equ10}
\end{equation}

After channel estimation, the channel vector between the $k$th user of cell $l$ and all of the RRUs of cell 1 can be expressed as a correlated fading channel with the Rayleigh fading part ${{\bm{\hat h}}_k}$. The main difference between the realistic channel model and the equivalent channel model is that the Rayleigh fading part of ${{\bm{\hat g}}_{l,k}}$ is not related to $l$ while that of ${{\bm{g}}_{l,k}}$ is statistically independent for any $l$.

By defining the following channel matrix ${{\bm{\hat G}}_l} = \left[ {\begin{array}{*{20}{c}}{{{{\bm{\hat g}}}_{l,1}}}& \cdots &{{{{\bm{\hat g}}}_{l,K}}}\end{array}} \right]$, the relationship between the estimated channel matrix ${{\bm{\hat G}}_l}$ and the transmit and receive signal can be expressed as
\begin{equation}
{{\bm{y}}_1} = {{\bm{\hat G}}_1}{{\bm{x}}_1} + \sum\limits_{l = 2}^L {{{{\bm{\hat G}}}_l}{{\bm{x}}_l}}  + \sum\limits_{l = 1}^L {{{{\bm{\tilde G}}}_l}{{\bm{x}}_l}}  + {{\bm{z}}_1},
\label{equ11}
\end{equation}
Using the minimum mean square error (MMSE) receiver, the sum-rate can be written as\cite{wang2013WCNC}
\begin{equation}
{ C} = {\log _2}\det \left( {\sum\limits_{l = 1}^L {{{{\bm{\hat{G} }}}_l}{\bm{\hat{G} }}_l^{\rm{H}}}  + {\bm{\Sigma }}} \right) - {\log _2}\det \left( {\sum\limits_{l = 2}^L {{{{\bm{\hat{G} }}}_l}{\bm{\hat{G} }}_l^{\rm{H}}}  + {\bm{\Sigma }}} \right),
\label{equ12}
\end{equation}
where
\begin{equation}
{\bm{\Sigma }} = \sum\limits_{l = 1}^L {\sum\limits_{k = 1}^K {\left( {{{\bm{R}}_{l,k}}{{\bm{\Lambda }}_{l,k}} - {{\bm{R}}_{l,k}}{{\bm{\Lambda }}_{l,k}}{\bm{Q}}_k^{ - 1}{{\bm{R}}_{l,k}}{{\bm{\Lambda }}_{l,k}}} \right)} }  + {\gamma _{{\rm{UL}}}}{{\bm{I}}_{MN}},
\label{equ13}
\end{equation}
and ${\gamma _{\rm UL}}$ is the Gaussian noise variance in the uplink channel.
As mentioned above, the equivalent model of multi-cell multi-user LSAS was given with imperfect CSI. Obviously, the above formulas are similar to that of MIMO channel capacity with multi-cell interference. However, it is more challenging to thoeretically analyze the capacity in LSAS, owing to the correlation of ${{\bm{\hat G}}_l}$. It has been proved in~\cite{wang2013WCNC} that as $MN \to \infty $, the sum-rate satisfies
\begin{equation}
{ C} - { C_{\inf }} \rightarrow 0.
\label{equ14}
\end{equation}
where
\begin{eqnarray}
\label{equ15}
& {C_{\inf }} = \sum\limits_{k = 1}^K {{{\log }_2}\left[ {1 + {\xi _{1,1,k}} - {\bm{\xi }}_{1,2:L,k}^{\rm{T}}{{\left( {{{{\bm{\Xi '}}}_k} + {{\bm{I}}_{L - 1}}} \right)}^{ - 1}}{{\bm{\xi }}_{2:L,1,k}}} \right]},\\
\label{equ16}
& {\xi _{i,l,k}} \buildrel \Delta \over = {\rm{Tr}}\left( {{\bm{Q}}_k^{ - 1}{{\bm{\Lambda }}_{i,k}}{{\bm{R}}_{i,k}}{{\bm{\Sigma }}^{ - 1}}{{\bm{R}}_{l,k}}{{\bm{\Lambda }}_{l,k}}} \right),\\
\label{equ17}
& {\bm{\Xi '}} = \left[ {\begin{array}{*{20}{c}}
{{\xi _{2,2,k}}}& \cdots &{{\xi _{2,L,k}}}\\
 \vdots & \ddots & \vdots \\
{{\xi _{L,2,k}}}& \cdots &{{\xi _{L,L,k}}}
\end{array}} \right],\\
\label{equ18}
& {{\bm{\xi }}_{1,2:L,k}} = {\left[ {\begin{array}{*{20}{c}}
{{\xi _{1,2,k}}}& \cdots &{{\xi _{1,L,k}}}
\end{array}} \right]^{\rm{T}}},\\
\label{equ19}
& {{\bm{\xi }}_{2:L,1,k}} = {\left[ {\begin{array}{*{20}{c}}
{{\xi _{2,1,k}}}& \cdots &{{\xi _{L,1,k}}}
\end{array}} \right]^{\rm{T}}}.
\end{eqnarray}
The above equations are general expressions for the capacity of multi-cell multi-user DAS, which covers those of massive MIMO. It is easy to verify that the expression of the sum-rate in~\cite{wang2013ICC} is a special case of (\ref{equ15}). In particular, when ${{\bm{R}}_{l,k}} = {{\bm{I}}_M}$ for all $l$ and $k$, $N=1$, and $M \to \infty $,  $C_{\inf }$ satisfies
\begin{equation}
{ C_{{\inf}}} = \sum\limits_{k = 1}^K {{{\log }_2}\left[ {1 + \frac{{\lambda _{1,1,k}^2}}{{\sum\limits_{l = 2}^L {\lambda _{l,1,k}^2}  + \frac{{\varepsilon  + {\gamma _{{\rm{UL}}}}}}{M}\left( {\sum\limits_{l = 1}^L {{\lambda _{l,1,k}}}  + {\gamma _{\rm{P}}}} \right)}}} \right]},
\label{equ20}
\end{equation}
where
\begin{equation}
\varepsilon  = \sum\limits_{k = 1}^K {\sum\limits_{l = 1}^L {{\lambda _{l,1,k}}} }  - \sum\limits_{k = 1}^K {\left( {\sum\limits_{l = 1}^L {\lambda _{l,1,k}^2} } \right){{\left( {\sum\limits_{l = 1}^L {{\lambda _{l,1,k}}}  + {\gamma _{\rm{P}}}} \right)}^{ - 1}}}.
\label{equ21}
\end{equation}
It can be seen that when $M \to \infty $,
\begin{equation}
{C_{{\rm{inf}}}} = \sum\limits_{k = 1}^K {{{\log }_2}\left( {1 + \frac{{\lambda _{1,1,k}^2}}{{\sum\limits_{l = 2}^L {\lambda _{l,1,k}^2} }}} \right)},
\label{equ22}
\end{equation}
which is the same as the sum-rate with the maximal ratio combining (MRC) receiver \cite{Marzetta_WC}.

The above theoretical results show that the capacity increases with the number of users, when the number of antennas at the BS is larger than the number of users. However, if orthogonal pilots are used in the uplink, pilot overhead increases linearly with the number of users, and then there exists a largest number of users served by the system. It can also be seen that LSAS is interference-limited and its performance is limited by the users using the same pilot sequence in other cells. The downlink capacity of large-scale DAS was studied with maximun ratio transmission (MRT) precoder according to the above model with pilot contamination in~\cite{Jiamin2015ICC}.  The results in ~\cite{Jiamin2015ICC, wang2013WCNC} demonstrated that under the same condition, the uplink capacity of large-scale distributed MIMO systems has 100$\%$ improvement than that of massive MIMO systems with MMSE receiver, while the improvement is 50$\%$ with MRT in the downlink.


\subsection{System-level spectral efficiency}
System-level spectral efficiency is a key parameter indicator of cellular mobile communication systems in industry, and is usually obtained by complex and time-consuming system simulation. In recent years, in order to demonstrate the impact of the system parameters, some researchers have studied the theoretical analysis of the system-level spectral efficiency.

There are two kinds of methods to study system-level spectral efficiency. The first one is using the stochastic geometry. As a representative paper, \cite{Andrews} derived the system-level spectral efficiency of multi-cell cellular systems. The main idea was that the spectral efficiency is obtained by Shannon formula according to signal to interference plus noise ratio (SINR), and then assuming that the BS locations are modeled as Poisson point process, the system-level spectral efficiency can be given by an elegant closed-form expression. The stochastic geometry model attracts more and more attention. For example, it has been applied to spectral efficiency analysis of cooperative BS systems \cite{Baccelli}, capacity analysis of heterogeneous network \cite{Fei}, capacity analysis of DAS \cite{YuWei2013ICC}, and capacity analysis of massive MIMO systems \cite{BaiArxiv}. However, in cooperative BSs or DAS, the closed expressions obtained by using Poisson point process is too complex to give an explicit relationship between system-level spectral efficiency and system parameters. Furthermore, when imperfect CSI (especially pilot contamination) is considered, due to the complicated expression of SINR, the analysis of system-level spectral efficiency has not been seen.

Another method is based on the model which has been widely used in the system-level simulation. The theoretical analysis has two steps: first, the locations of the users in the system are fixed, and the ergodic spectral efficiency is obtained by averaging the spectral efficiency over the small-scale fading. Obviously, the ergodic spectral efficiency is a function of the users' location. After that, assuming that the BS locations are fixed  and all the users are uniformly distributed in every cell, we obtain the average of the ergodic spectral efficiency. The closed expressions for both distributed and centralized antenna systems were firstly derived in \cite{Wang2008ICC}. In \cite{yang_performance_2014}, the average spectral efficiency of DAS was further obtained with all RRUs on a circle. According to asymptotic upper and lower bounds of system-level spectral efficiency for non-cooperative multi-cell multi-user cellular systems, the relationship between system-level spectral efficiency and system parameters was derived (including bandwidth, the number of users and RRUs) in \cite{aggarwal_design_2013}. In \cite{Xin_2015}, an approximate closed-form expression of average spectral efficiency for multi-cell multi-user massive MIMO systems was given when pilot reuse was considered.

\subsection{Topics for Future Research}
When pilot contamination is considered, spectral efficiency analysis should be further studied for large-scale DAS with zero forcing (ZF) or regularized ZF (RZF) precoding. Recently, the impact of non-ideal hardware on system capacity attracts attention\cite{Bjornson}\cite{Gustavsson}. Moreover, it was found that non-reciprocity between uplink and downlink has a substantial impact on spectral efficiency for time division duplex (TDD) systems.  Finally, considering imperfect CSI, asymptotic analysis of system-level spectral efficiency is also a challenging research direction.

\section{CSI acquisition in LSAS}
In LSAS, with an increasing number of BS antennas and spatial-division users, CSI acquisition becomes a bottleneck in system implementation. When orthogonal pilots are employed, pilot overhead increases linearly with the total number of antennas of the users or the total number of antennas at BS for uplink or downlink, respectively. To achieve the capacity of broadcast channel, CSI of downlink should be known at BS. However, since the number of BS antennas may be much larger than the total number of users' antennas, downlink CSI acquisition becomes an obstacle for LSAS. For TDD systems, during the coherence time, due to the reciprocity of the uplink and downlink channels, the estimation of uplink channels at BS can be utilized to design downlink precoding. That means the overhead of both downlink pilots and the CSI feedback can be avoided for TDD systems. While for frequency division duplex (FDD) systems, the lack of the reciprocity becomes a main challenge.

CSI acquisition in LSAS is experiencing  facing the following problems. First, the pilot overhead still linearly increases with the total number of users' antennas, even for TDD systems. How to reduce the pilot overhead or utilize pilot resource efficiently should be studied. Second, under the limited pilot resource, how to improve the accuracy of channel estimation is also important. Finally, although for TDD system, the uplink/downlink wireless channels are reciprocal, the overall channels are nonreciprocal due to the mismatch of transceiver radio frequency (RF) circuits. In the following, we will give the detailed investigation on the pilot design, channel estimation methods, and reciprocity calibration for LSAS.

\subsection{Pilot design}
Since CSI plays an important role in both transmitter and receiver, reference signal design has always been the key technology for mobile communication systems. According to the functions of reference signals, they can be divided into CSI reference signal (CSI-RS) and demodulation reference signal (DM-RS). CSI-RS is usually sent omnidirectionally, which can be used to obtain the channel quality indicator, or the statistical channel information, etc.. CSI-RS is usually sparse in time and frequency domain. DM-RS are mainly served as data demodulation. To reduce the overhead, precoded pilots are usually adopted for DM-RS.

Pilot design is usually divided into two kinds: orthogonal design and non-orthogonal design. Orthogonal pilot sequences has been widely used in wireless communication systems. The idea of orthogonal design includes time division multiplexing, frequency division multiplexing, code division multiplexing and their combinations. These techniques have been applied in 3GPP LTE standard. The advantage of orthogonal design is that it can achieve optimal performance with least square channel estimation. However, the disadvantage is that the overhead is large for LSAS, especially for multi-cell multi-user LSAS. Hence, there is an urgent request to reduce the pilot overhead. Nonorthogonal pilot sequences are often  adopted to reduce the pilot overhead. For LSAS, two solutions have been proposed. One uses the idea of superposition of pilot symbols and data symbols, and another solution is pilot reuse. The former will introduce extra interference between pilot and data, and the later will suffer from pilot contamination.

In \cite{Fernandes_2013}, a multi-cell time-shifted pilot scheme was presented, and the interference between pilots and data can be suppressed by interference cancellation. \cite{Fernandes_2013} showed that with this method, pilot contamination can be mitigated when the number of BS antennas approaches infinity. However, when the number of spatial-division users is large, for time-shifted pilot systems, to achieve the same performance as the synchronized pilot systems, there is need for  more antennas in the BS. In \cite{Hua}, the semi-orthogonal pilot design was proposed to allow simultaneous data and pilot transmission by using the successive interference cancellation. Simulation and theoretical analysis showed that the method can greatly reduce the pilot overhead and then improve the spectral efficiency. The idea behind both \cite{Fernandes_2013} and \cite{Hua} is the superposition of  pilot and data.

Pilot reuse \cite{Marzetta_WC} has been studied since massive MIMO was firstly proposed. In \cite{Shi2015Pilot}, with long-term channel information, such as large-scale fading, three low-complexity pilot scheduling approaches were further proposed to maximize the achievable sum rate, including the greedy algorithm, the tabu search (TS) algorithm and the greedy TS algorithm. The research on massive MIMO channel model showed that there are  sparse characteristics in the spatial domain and time domain when both the number of antennas and the bandwidth greatly increase. Pilot assignment using the sparsity of the massive MIMO channels can mitigate pilot contamination effectively. In \cite{YouLiTWC}, it has been proved that the mean square error of channel estimation can be minimized, provided that the angle of arrival intervals of users are non-overlapping. Therefore, pilot reuse in spatial correlated massive MIMO channels is feasible when channels are sparse in the angle domain. Moreover, the second-order statistics of channels can be exploited to improve the performance of pilot assignment \cite{Shi2015Pilot,YouLiTWC,yin_coordinated_2013}, and therefore, pilot contamination can be mitigated. In the time domain, the sparsity of the wideband MIMO channels can also be used to suppress pilot contamination \cite{chen_pilot_2014}\cite{gao_structured_2014}.

For large-scale DAS, the distributed MIMO channels are sparse in the power domain \cite{wang2015Large}. We can also use this property of distributed MIMO to do pilot assignment. With the users' locations, an inter-user interference (IUI) matrix can be constructed to quantify the inter-user pilot interference.  From this point of view, pilot reuse and frequency allocation are similar. Some methods used in frequency allocation, such as fractional frequency reuse or the advanced graph coloring algorithm, can be employed to reduce the pilot contamination in distributed MIMO systems \cite{Yangyaoqing}\cite{Atzeni}.

For CSI acquisition in massive MIMO systems, there are more challenges in FDD than in TDD. Assuming that the BS and the users share a common set of training signals,  open-loop and closed-loop training frameworks were proposed in \cite{Choi_2014}. In open-loop training, the BS transmits training signals in a round-robin manner, so that the receiver can estimate the current channel using spatial or temporal correlations as well as previous channel estimations. In closed-loop mode, the user feeds back the sequence number of the best training signal which is selected from the previously received training signals. Then, with the index feedback from the users, the BS determines the training signal to be sent out.

\subsection{Channel estimation methods}
Exploiting the sparsity of massive MIMO channel is an efficient way to improve the performance of channel estimation. Based on a physically motivated channel model, a critical relationship for massive MIMO orthogonal frequency division multiplexing (OFDM) has been revealed between the space-frequency domain channel covariance matrix (SFCCM) and the channel power angle-delay spectrum \cite{ You2015Channel }. When the number of antennas at BS is sufficiently large, the SFCCM can be well approximated by \cite{ You2015Channel}
\begin{equation}\label{eq_youli}
{{\bm{R}}_k} \approx \left( {{{\bm{F}}_{{N_{\rm{c}}} \times {N_{\rm{g}}}}} \otimes {{\bm{V}}_M}} \right){\rm{diag}}\left[ {{\rm{vec}}\left( {{{\bm{\Omega }}_k}} \right)} \right]{\left( {{{\bm{F}}_{{N_{\rm{c}}} \times {N_{\rm{g}}}}} \otimes {{\bm{V}}_M}} \right)^{\rm{H}}},
\end{equation}
where $N_{\rm c}$ is the number of subcarriers, $N_{\rm g}$ is the length of the guard interval, ${{\bm{F}}_{{N_{\rm{c}}} \times {N_{\rm{g}}}}}$
denotes the matrix composed of the first $N_{\rm g}$ columns of $N_{\rm c}$ dimensional unitary discrete Fourier transform (DFT) matrix,
${{\bm{V}}_M}$ is a matrix containing the spatial information, and ${{\bm{\Omega }}_k}$ is referred to the angle-delay domain channel power matrix, which describes the sparsity of wireless channels in the angle-delay domain. Therefore, the eigenvectors of the SFCCMs for different users tend to be the same in the asymptotically large array regime, which shows that massive MIMO-OFDM channels can be asymptotically decorrelated by the fixed space-frequency domain statistical eigendirections, while the eigenvalues depend on the corresponding channel power angle-delay spectrum. Furthermore, with a presented equivalent channel model, the sum of mean square error of channel estimation can be minimized if the channel power distributions of different users in the angle-delay domain can be made non-overlapping by proper pilot phase shift scheduling.

Since channels in wideband massive MIMO have sparsity in both the angular {\cite{ Adeogun2014Channel}} and delay domain {\cite {chen_pilot_2014}}, a parametric channel model can be utilized to represent this type of channels. Then, subspace channel estimation methods {\cite{simeone_pilot-based_2004, huang2006low}} can be applied to parameterized channel estimation. In {\cite{zhu_doa_2013}} , more accurate channel estimates could be obtained through direction of arrival and direction of departure estimation of resolvable paths.

Compress sensing, as another effective way to estimate sparse channels, can achieve more accurate channel estimation with lower pilot overhead. In {\cite{ qi2015sparse }}, with the inherent sparsity of wireless channels, the channel estimation was modeled as a joint sparse recovery problem, which can be solved by an improved algorithm named block optimized orthogonal matching pursuit. In {\cite{Masood2015}}, a distributed Bayesian method, which is based on the support agnostic Bayesian matching pursuit algorithm {\cite {Masood2013}}, has been developed to improve channel estimates in massive MIMO. This approach essentially exploits the channel sparsity and common support properties to estimate sparse channels with a small number of pilots.

The problem of pilot contamination is very similar to the channel estimation problem in code division multiple access (CDMA) systems. Blind channel estimation which has been extensively studied in CDMA system by many researchers was reconsidered to solve the problem of the pilot overhead for LSAS. In TDD systems, with spatial asymptotic orthogonality of channels in massive MIMO systems, Ngo et al. proposed a blind channel estimation algorithm based on eigenvalue decomposition\cite{NgoICASSP}. Using only a few uplink pilots, it can remove the ambiguity of the blind channel estimation. Based on subspace projection, a blind channel estimation method without pilot symbols was presented to further overcome pilot contamination \cite{Muller}. However, the computational complexity becomes a major obstacle for the practical implementation.

In addition, data-aided channel estimation is also a traditional method to improve the performance of channel estimation. The performance can be further enhanced by using iterative receiver with joint data detection and channel estimation \cite{MaTSP2015}. Nevertheless, due to the non-orthogonality of the data symbols, the performance improvement can be obtained only when the data length is  long enough. Correspondingly, in return, its computational complexity increases significantly with the number of users and the data length.

By exploiting the sparsity in the time domain, a joint design of channel estimation and  pilot assignment was proposed to mitigate pilot contamination as well as improve the performance in \cite{chen_pilot_2014}. The main idea behind this method is that with the orthogonal property of different users in the time-delay domain, pilot contamination is randomized by pilot assignment in the different time slots. Then, the power delay profile (PDP) is estimated via multipath time-delay estimation and multipath component extraction. In the ideal case, the estimated PDP approximates to that without pilot contamination. Finally, pilot contamination can be eliminated by the estimated PDP.

In an FDD system, feedback is a typical way to improve the accuracy of channel estimation. In \cite{RaoTSP2014}, a distributed compressive CSI estimation scheme was proposed, where the compressed measurements were observed at the users locally, while the CSI recovery was performed at the BS jointly. Considering the spatial sparsity and slow-varying characteristics, a distributed sparsity adaptive matching pursuit algorithm was proposed in \cite{GaoTSP2015}. By exploiting the spatially common sparsity of massive MIMO channels during multiple time blocks, a closed-loop channel tracking scheme was also proposed to track the channels \cite{GaoTSP2015}.

In large-scale distributed MIMO, both the small-scale fading coefficients (SSFCs) and the large-scale fading coefficients (LSFCs) play important roles. Usually, the estimation of LSFCs has been largely neglected, assuming somehow perfectly known prior to SSFCs estimation. In \cite{chen_composite_2013}, a channel estimation algorithm was proposed to obtained the LSFCs by taking advantage of the spatial samples and the channel hardening effect.

\subsection{Reciprocity calibration for TDD systems}
In the practical wireless communication systems, the baseband channel matrix is composed of the wireless propagation channel and the coefficients of transceiver RF circuits. For TDD systems, although the wireless propagation channels are reciprocal, the RF circuits usually include antennas, mixers, filters, analog to digital converters, power amplifiers, etc. which are usually not symmetric for transmitter and receiver. This non-symmetric of the transceiver RF (also called as RF mismatches) makes the whole channel matrix  non-reciprocal.

Considering the RF mismatches, the uplink and downlink channels are modeled by
\begin{eqnarray}
\label{equ23}
& {{\bm{G}}_{{\rm{UL}}}} = {{\bm{C}}_{{\rm{BS}},{\rm{r}}}}{{\bm{H}}^{\rm{T}}}{{\bm{C}}_{{\rm{UE}},{\rm{t}}}},\\
\label{equ24}
& {{\bm{G}}_{{\rm{DL}}}} = {{\bm{C}}_{{\rm{UE}},{\rm{r}}}}{\bm{H}}{{\bm{C}}_{{\rm{BS}},{\rm{t}}}},
\end{eqnarray}
where $\bm{H}$ represents the wireless channel matrix, ${{\bm{C}}_{{\rm{BS}},{\rm{t}}}}$ and ${{\bm{C}}_{{\rm{BS}},r}}$ denote the transmit and receive RF matrices of the BS, respectively, and ${{\bm{C}}_{{\rm{UE}},{\rm{t}}}}$ and ${{\bm{C}}_{{\rm{UE}},r}}$ denote the transmit and receive RF matrices of the user equipment (UE), respectively. All of these RF matrices are modeled as a diagonal matrix. Each diagonal can be modeled as
a random variable with log-normal distributed amplitude and uniform distributed phase.

Using the transpose of the uplink channel as the downlink channel, the ZF precoding matrix is written as
\begin{equation}
{\bm{W}} = {\bm{G}}_{{\rm{UL}}}^{\rm{*}}{\left( {{\bm{G}}_{{\rm{UL}}}^{\rm{T}}{\bm{G}}_{{\rm{UL}}}^{\rm{*}}} \right)^{ - 1}}.
\label{equ25}
\end{equation}
It can be seen that ${{\bm{G}}_{{\rm{DL}}}}{\bm{W}}$ is not a diagonal matrix and there exists IUI. If we use the following precoding matrix
\begin{equation}
{\bm{W}} = {\bm{C}}_{{\rm{BS,t}}}^{ - 1}{{\bm{C}}_{{\rm{BS,r}}}}{\bm{G}}_{{\rm{UL}}}^{\rm{*}}{\left( {{\bm{G}}_{{\rm{UL}}}^{\rm{T}}{\bm{G}}_{{\rm{UL}}}^{\rm{*}}} \right)^{ - 1}},
\label{equ26}
\end{equation}
${{\bm{G}}_{{\rm{DL}}}}{\bm{W}}$ becomes a diagonal matrix which means the IUI is avoided. Therefore, ${\bm{C}}_{{\rm{BS,t}}}^{ - 1}{{\bm{C}}_{{\rm{BS,r}}}}$ is called as the calibration matrix.

The impact of the RF mismatches on the massive MIMO systems was analyzed in \cite{Wei2015Mutual,Zhangwence,Wei2016Impact}. Considering the RF mismatches, the lower bound of the ergodic sum-rate for multi-user massive MIMO can be given by \cite{Wei2016Impact}
\[{\cal E}\left( {{R^{{\rm{mis}}}}} \right) \ge R_{{\rm{LB}}}^{{\rm{perfect}}} - \Delta R_{{\rm{BS}}}^{{\rm{mis}}} - \Delta R_{{\rm{UE}}}^{{\rm{mis}}}\]
where
\[R_{{\rm{LB}}}^{{\rm{perfect}}} = K \cdot \left[ {\log \left( \rho  \right) + \log \left( {\frac{{M - K}}{K}} \right)} \right]\]
\[\Delta R_{{\rm{BS}}}^{{\rm{mis}}} = K \cdot \left\{ {\log \left( \rho  \right) + \log \left( {\frac{{{\lambda _1}}}{{{\lambda _2}}}} \right) + \log \left[ {\frac{{MK}}{{\left( {M - K} \right)\left( {K - 1} \right)}}} \right]} \right\}\]
\[{\lambda _1} = {\rm{sin}}{{\rm{c}}^2}\left( {{\theta _{{\rm{BS}},{\rm{r}}}}} \right) \cdot {\rm{sin}}{{\rm{c}}^2}\left( {{\theta _{{\rm{BS}},{\rm{t}}}}} \right)\]
\[{\lambda _2} = {e^{2\delta _{{\rm{BS}},{\rm{t}}}^2}} + {e^{2\delta _{{\rm{BS}},{\rm{r}}}^2}} - 2{e^{\delta _{{\rm{BS}},{\rm{t}}}^2/2 + \delta _{{\rm{BS}},{\rm{r}}}^2/2}} \cdot {\rm{sinc}}\left( {{\theta _{{\rm{BS}},{\rm{r}}}}} \right) \cdot {\rm{sinc}}\left( {{\theta _{{\rm{BS}},{\rm{t}}}}} \right)\]
\[\Delta R_{{\rm{UE}}}^{{\rm{mis}}} = K\left( {\log e \cdot 2\delta _{{\rm{UE}},{\rm{t}}}^2} \right),\]
$\rho$ denotes the signal-to-noise ratio (SNR), $\delta^2$ denotes the variance of the amplitude mismatch, and $\theta$ denotes the range of the phase mismatch for corresponding transceiver (the subscripts 't' and 'r' denote transmitter and receiver, respectively).

\begin{figure}[!t]
\centering
\begin{minipage}[c]{0.48\textwidth}
\centering
\includegraphics[scale=.5]{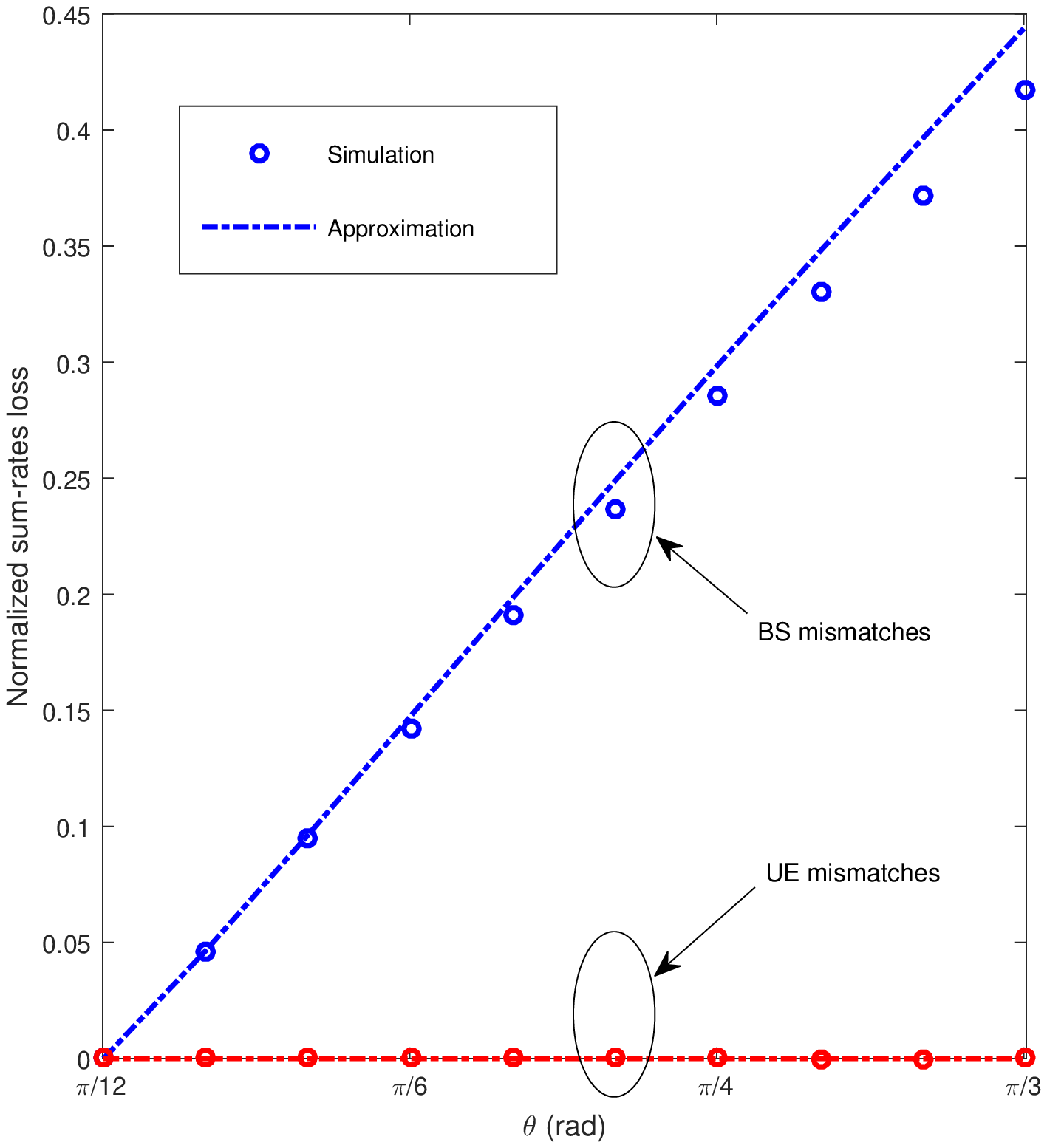}
\end{minipage}
\hspace{0.02\textwidth}
\begin{minipage}[c]{0.48\textwidth}
\centering
\includegraphics[scale=.5]{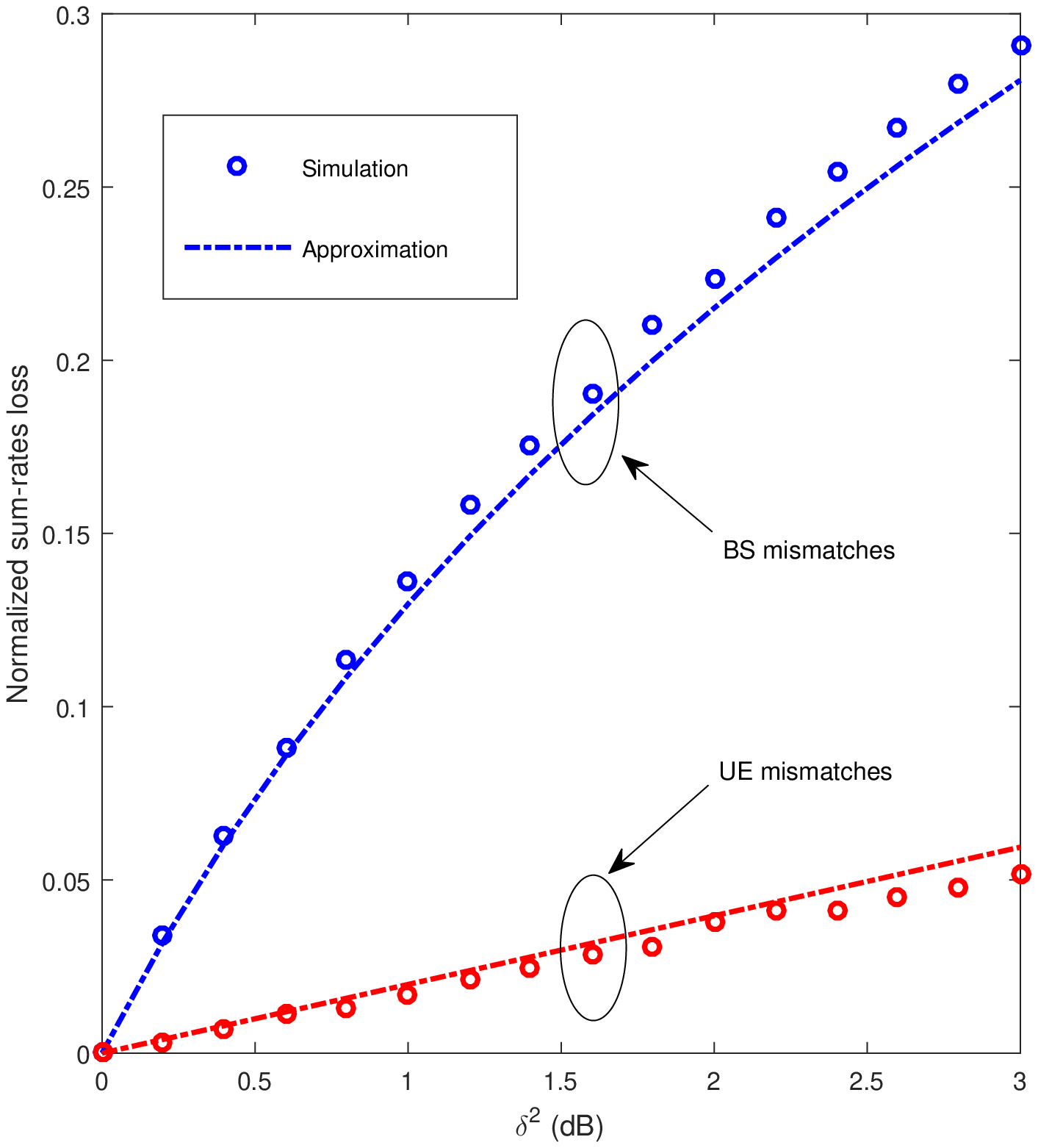}
\end{minipage}\\[3mm]
\begin{minipage}[t]{0.48\textwidth}
\centering
\caption{Normalized sum-rates loss versus the phase range of RF mismatches. The transmit SNR is set to be 10dB.}
\label{figphase}
\end{minipage}
\hspace{0.02\textwidth}
\begin{minipage}[t]{0.48\textwidth}
\centering
\caption{Normalized sum-rates loss versus the amplitude variance of RF mismatches. The transmit SNR is set to be 10dB.}
\label{figamp}
\end{minipage}
\end{figure}

From Figure~\ref{figphase} it can be seen that the RF phase mismatch of BS will introduce more than 40$\%$  performance degradation for ${{\theta _{{\rm{BS}},{\rm{t}}}}} = {{\theta _{{\rm{BS}},{\rm{r}}}}} = \pi/3$. However, the phase mismatch of UE does not degrade the system throughput. The simulation results are consistent with the theoretical analysis.   From Figure~\ref{figamp} it can be seen that for the BS, the amplitude mismatch also introduces severe performance degradation. For example, the performance loss is about 28$\%$ for ${{\delta^2 _{{\rm{BS}},{\rm{t}}}}} = {{\delta^2 _{{\rm{BS}},{\rm{r}}}}} = 3$ dB. While for the UE, the performance decreases slightly if the variance of the amplitude mismatch is small.

Therefore, it is very important to perform the reciprocity calibration for the implementation of massive MIMO systems and large-scale distributed MIMO systems, especially for the BS. There are two types of calibration methods, one is called hardware calibration method, and another is named signal-space calibration method.

The idea of the hardware calibration method is to use the couplers and multiple switches to connect the transmit circuits of each antenna with the receive circuits of the rest of the antennas \cite{Nishimori}. Then, the self-transmitted calibration signals are used to perform the RF chains measurement. The hardware calibration method is preferred for massive MIMO systems due to the advantages of fastness and simplicity. However, this method requires a huge hardware cost, which is a heavy overhead for massive MIMO systems. Besides, this method may not be used for large-scale distributed MIMO systems, since the antennas of different RRUs are geographically separated. To avoid extra hardware circuits, the signal-space calibration method was proposed, where the calibration coefficients are calculated by the calibration signals. Then, this method is also named over-the-air (OTA) calibration method.

OTA calibration can be classified as partial calibration and full calibration. The partial calibration was proposed to just calibrate the RF mismatches at the BS, for example the calibration matrix in (\ref{equ26}). While, the full calibration compensates for the RF mismatches at both the BS and the UEs. For full calibration method, we need to  exchange the calibration signals between BS and UEs. Then, the BS computes the calibration coefficients according to the received uplink calibration signals and the downlink calibration signals feedback from the UEs. In \cite{kaltenberger2010relative}, total least squares (TLS) based calibration method was proposed to achieve the optimal performance. But, in practice, the UEs are desired to be excluded from the calibration procedure, and the feedback will become very heavy with the increment of the antennas in massive MIMO systems.

Actually, we only need to perform calibration at the BS, since the RF mismatches of the UEs have a negligible impact on the system performance \cite{Wei2015Mutual}. In \cite{Shepard}, Argos method was proposed to perform calibration for an experimental massive MIMO system. It relies on the sequential transmission and reception of pilots between the reference antenna and the rest of the antennas. For distributed MIMO systems, the calibration coefficients are obtained by exchanging calibration signals between the reference RRU and the other RRUs \cite{Rogalin}. Nevertheless, the system performance will degrade significantly, when the SNR between the reference antenna and the other antennas is very small \cite{Shepard}\cite{Rogalin}. Thus, to improve the calibration performance, LS calibration was presented in \cite{RogalinTWC}. Not only relying on the reference antenna, the LS method utilizes the received calibration signals of all antennas to obtain the better performance. Theoretically, in \cite{Wei2015TDD} the LS method has been proved to be equivalent to the TLS method, since the LS method can be regarded as the extension of the TLS method \cite{kaltenberger2010relative} to the case of self-calibration. To avoid the eigenvalue decomposition required by the LS method, an iterative coordinate decent method was proposed to achieve the performance of the LS method yet with low computation complexity\cite{Wei2015TDD}.

\subsection{Topics for Future Research}
At present, CSI acquisition is regarded as the key technology to implement LSAS whether in academia or industry. Exploiting the channel characteristic is often the preferable way to improve the performance of channel estimation. The parametric modeling and the sparse nature of LSAS are beneficial for channel parameters estimation and the pilot overhead reduction. For large-scale distributed MIMO systems, frequency and time synchronization among RRUs are basic prerequisites for achieving the joint processing gain. In addition, calibration methods should be further studied. Although some calibration methods, such as TLS and LS, could achieve good performance in TDD systems, the computational complexity is still very high. For FDD systems, it is always a hot topic to reduce the feedback overhead for MU-MIMO systems. The reciprocity of statistical CSI (e.g., the correlation matrix of antennas) in FDD systems has been looked as an efficient way to reduce the feedback overhead. However, the necessity of calibration for the statistical CSI should be validated by experimental systems.

\section{Transmission Methods of Multi-user LSAS}
As known, for uplink multi-user MIMO systems, the capacity-approaching receiver is  maximum a posteriori  detection. For downlink transmission, the capacity can be achieved by dirty-paper coding with the downlink channel information. However, with the increasing number of antennas and users, it is difficult to implement optimal transmission. The theoretical results of\cite{Marzetta_WC} and \cite{wang2013WCNC} demonstrated that for both massive MIMO and large-scale distributed MIMO systems, when the number of antennas tends to infinity, the channel capacity can be approached by low-complexity MRT in the downlink and MRC in the uplink. Furthermore, the results also showed that the channel capacity can be achieved by RZF precoding or linnear MMSE detection with not-so-large number of antennas \cite{Hoydis}. That is, for LSAS, the capacity-approaching transceiver is more practical.

Nevertheless, due to the capabilities of the current technologies, the scale of antennas can not be very large. When RZF precoding or linnear MMSE detection is employed, the computational complexity of the matrix inversion is still very large for a large number of users. Currently, there are two approaches to overcome the obstacle. One is to find a low-complexity matrix inversion method. Another way is to reduce the complexity of the transceiver by making use of the sparsity or the statistical property of the channel. In fact, due to the duality of the uplink and downlink, the mechanisms of both uplink and downlink transmission methods are very similar. For instance, in  \cite{Kammoun2014Linear} and \cite{HuangYuqi},  a polynomial expansion based precoding and a polynomial expansion based iterative receiver were proposed for massive MIMO to simplify matrix inversion, respectively. In order to exploit  the sparsity of LSAS, belief propagation (BP) was extended to downlink precoding \cite{Wen2014Message} and iterative detection \cite{wang2015Large}.

In the rest of this section, we will firstly introduce the space division multiple access (SDMA) transmission methods, and then present a joint transmission method for large-scale DAS. Finally, we focus on capacity-approaching receiver for LSAS.

\subsection{SDMA based on statistical CSI}
For the multi-user downlink transmission, in order to achieve the capacity of multi-user broadcast channel, it is usually assumed that downlink CSI  is known at the transmitter. Therefore, under this condition, efficient information transmission relies on the design of multi-user signals at the transmitter. However, different from tranditional MU-MIMO, there are two main difficulties to be solved when we design the downlink transmitter of multi-user massive MIMO. As we have discussed, first one is the complexity of CSI acquisition, and second on is the complexity of the spatial division for large-scale users. In order to solve both of the problems, multi-user transmission methods with statistical CSI have been proposed, including beam division multiple access (BDMA) \cite{sunTCOM} and joint spatial division and multiplexing (JSDM) \cite{NamTIT2013}.

The main idea of JSDM is a two-stage precoding scheme by using statistical CSI. First of all, the users in service are divided into groups, and each group of users have similar transmit correlation matrix. Then, the first-stage precoding is applied to each group of users semi-statically. Subsequently, the second-stage precoding is applied to the equivalent channel with the reduced dimension. It can be seen that the performance of the systems is largely dependent on user grouping.

For wideband massive MIMO systems, a beam domain channel model was proposed for theoretical analysis \cite{sunTCOM}. Based on the channel model, an upper bound of achievable ergodic sum-rate was derived. \cite{sunTCOM} showed that SDMA in the beam domain is optimal to maximize the upper bound of the ergodic sum-rate. That means by selecting users within non-overlapping beams, the MU-MIMO channels can be equivalently decomposed into multiple single-user MIMO channels. Then, a BDMA transmission scheme was proposed where multiple users are served simultaneously via different beams. With BDMA, both the complexity of transceiver and the overhead of channel estimation can  be reduced significantly.

\subsection{Downlink transmission in large-scale DAS}
In large-scale DAS, the sparsity in the power domain can be utilized to reduce the complexity of RZF precoding. In \cite{Wen2014Message}, RZF beamforming (RZFBF) was realized in a distributed manner by using BP, so that it can avoid large dimensional matrix inversion. Using Bayesian inference and Gaussian approximation, BP-RZFBF was further developed. By exploiting the channel hardening property in large-scale MIMO systems, the AMP-RZFBF based on spatial channel covariance information (CCoI), called CCoI-aided AMP-RZFBF, was also proposed, which results in much simpler implementations in terms of computation.

\begin{figure}[!t]
\centering
\includegraphics[scale=.5, angle=-90]{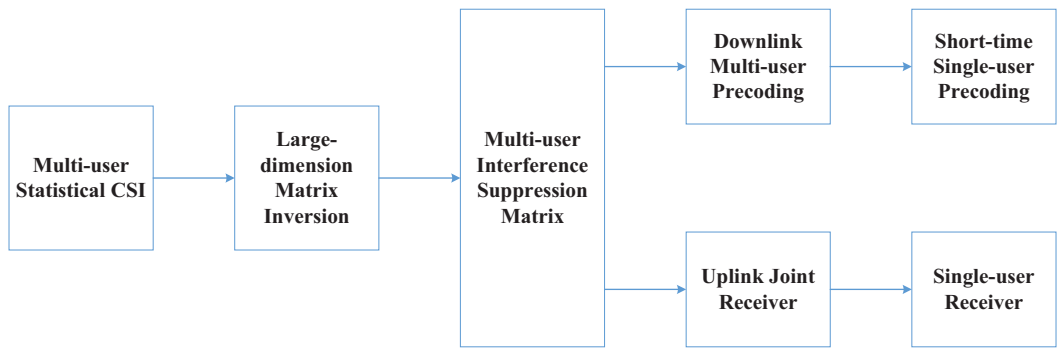}
\caption{Transmitter and Receiver Design for Large-scale distributed MIMO.}
\label{fig3}
\end{figure}

In TDD large-scale distributed MIMO-OFDM systems, a low-complexity transmission method was presented by a joint design of uplink and downlink. The idea is shown in Figure~\ref{fig3}. Firstly, we obtain the wideband statistical CSI matrix by uplink channel estimation. After TDD calibration and sparsification, we compute the precoding matrix by using the large-scale sparse matrix inversion. Then, we obtain the IUI suppression matrix. Finally, we perform uplink precoding and downlink precoding (called short-term single-user precoding) for every served user, according to the equivalent channel with the reduced dimension, i.e., the composite channel consisting of each user's practical channel and the interference suppression matrix. During the uplink transmission, each user sends its precoded signals, and at the BS, the received signals are passed to the interference suppression matrix, and then to the single-user detection. For the downlink, single-user precoding is performed for each user before IUI suppression precoding. Notice that the method could reduce the system complexity greatly, since we only compute the inverse of large dimensional statistical CSI matrix by using its sparsity.

\subsection{Low-complexity receivers}
The capacity-approaching receiver technology has always  attracted much attention in multi-antenna systems. For MIMO receiver, the trade-off between computational complexity and performance is also a hot-topic. In particular, since LSAS is interference-limited, it is a quite challenging task to design a receiver which achieves its capacity under the interference channels.

For LSASs, the following difficulties are encountered in the design of a receiver, which covers imperfect CSI, inter-cell interference and high-complexity of the multi-user detection.
In \cite{Narasimhan} a joint channel estimation and data detection algorithm with message passing was proposed by exploiting the channel hardening effect of massive MIMO. To overcome the complexity of the detection, recently, some researchers have studied large dimensional MIMO detection include simplified matrix inversion \cite{HuangYuqi}\cite{Dai2014Low-complexity} and sparsity-based detection \cite{wang2015Large}\cite{Fadlallah2014New}.

In \cite{HuangYuqi}, a turbo receiver has been proposed for massive MIMO. The iterative receiver consists of channel estimation, interference estimation, low-complexity soft-input soft-output (SISO) detection and decoding . Firstly, the channel parameters and preliminary statistical interference characteristics are estimated, and then linear prefiltering is peformed. Next, a low complexity SISO detector based on singular value decomposition is applied to the reduced-dimension channel matrix. To further reduce the detection complexity, a SISO detector based on polynomial expansion is presented to avoid matrix inversion. The results showed that due to the effect of large-scale antennas, the performance of the proposed turbo receiver is very close to the theoretical channel capacity performance.

\subsection{Topics for Future Research}
There are still many challenges on the wireless transmission of LSAS. Firstly, for high mobile speed users, it remains unknown that whether statistical CSI can be used to reduce the complexity of the implementation. The results in \cite{CaoJuan} indicated that large-scale MIMO has the ability to enhance the transmission rate of the high-speed users. However, the effectiveness of statistical beamforming and channel estimation of LSAS for high-mobility users should be further studied. Secondly, the wireless transmission methods of LSAS should be further studied under imperfect CSI, including channel estimation error and hardware impairment, such as I/Q imbalance \cite{WenceICC2015}, reciprocity mismatch \cite{HanTCOM}, a low-resolution analog-to-digital converter \cite{Fan}\cite{ZhangJiangTao}, etc.. Finally, the receiver of the BS has been paid a lots of attention. However, for downlink transmission, with BDMA or other SDMA using statistical CSI, due to the residual interference, the user equipment is usually confronted with the interference channel. How to improve the performance of the receiver of the UE is also a challenge.

\section{Resource allocation in LSAS}
In LSAS, in order to further improve the efficiency of radio resource in the network, it is necessary to jointly optimize the allocation of space-time resources, power resources and user groups, etc., which brings new challenges to radio resource management. Hence, a low-complexity and high-performance resource allocation strategy is essential for the implementation of LSAS. Many researchers have  devoted to the time-frequency resource allocation for MIMO-OFDM systems \cite{ZhuHuiling01,ZhuHuiling02,ZhuHuiling03}. Nevertheless, this paper mainly focuses on the spatial resource allocation strategies and the RRUs grouping strategies in LSAS. In the following, we will discuss  the spatial resource allocation strategies for massive MIMO and large-scale distributed MIMO, respectively.

The SDMA with statistical CSI can reduce the implementation complexity in massive MIMO systems. However, its performance is largely dependent on user grouping. A greedy algorithm based on statistical CSI has been applied to do user grouping by maximizing the sum capacity \cite{sunTCOM}. As mentioned above, those users who have similar eigenvectors of their channel covariance matrices should first be assigned into a same cluster for JSDM. Thus, the performance of the grouping method depends on IUI. An K-Means clustering method using the chordal distance as its grouping metric was proposed in \cite{Nam2014Joint}. In \cite{YueGuosen}, an improved grouping method using the weighted likelihood similarity measure and subspace projection based similarity measure has been proposed to provide a better system throughput.

In large-scale distributed MIMO systems, the clustering of RRUs and users is regarded as the key technology to reduce the system complexity. The clustering algorithms can roughly be classified into three kinds: the first one is the user-centric  clustering, the second one is the dynamic clustering and the last one is the quasi-static interlaced clustering.

The user-centric wireless communication was first proposed in the development of 4G \cite{Xu2007maximum}, and it is still a hot topic for 5G mobile communication at present. A large-scale DAS where each user chooses a few surrounding RRUs as its virtual cell was analyzed and the optimal size of the virtual cell was given in \cite{Dai2014An}. Under the capacity constraint of the inter-RRU backhaul, a method combining the user-centric dynamic clustering, the user scheduling and beamforming was proposed by maximizing the weighted sum rate with the generalized weighted mean square error \cite{Dai2014Sparse}.

Statistical CSI can also be used in the dynamic clustering \cite{Liu2009An}. In \cite{Zhangyingjun}, using the sparsity of the channels, the channel matrix was transformed into a bilateral block diagonal matrix, with which the dynamic clustering and the parallel baseband processing were realized.

A quasi-static overlapping for BS clustering was presented in \cite{Ratnam2015Capacity}. All the RRUs in the system are divided into disjoint clusters, and different clustering methods lead to different cluster patterns between which there exists an overlap. In each cluster pattern, all the users and RRUs employ the same frequency resources, while orthogonal time-frequency resources are allocated in different patterns. Combined with a power allocation method, the clustering method can significantly improve the throughput of the whole cell as well as its boundary.

There has always been much interest in resource allocation in wireless communication. Moreover, it is of great value to design resource allocation and multi-user scheduling methods with lower complexity in LSAS.

\section{Conclusions}
In this paper, the state-of-the-art research progress on the wireless transmission theory and technology in LSAS have been summarized, which covers the theoretical analysis of the spectral efficiency, CSI acquisition, uplink and downlink wireless communication, and resource allocation. Besides, this paper also presented potential research topics. Although there have already been some experimental large-scale MIMO systems, and the 3GPP standard organization is also promoting the evolution of relevant technology, a large number of theoretical and engineering problems should be addressed.

\ifCLASSOPTIONcaptionsoff
  \newpage
\fi

\end{document}